\documentclass[12pt,preprint]{aastex}
\usepackage{graphicx}
\begin{document}
\def\la{\mathrel{\hbox{\rlap{\hbox{\lower4pt\hbox{$\sim$}}}\hbox{$<$}}}}
\def\ga{\mathrel{\hbox{\rlap{\hbox{\lower4pt\hbox{$\sim$}}}\hbox{$>$}}}}
\def\lam{$\lambda$}
\def\kms{km~s$^{-1}$}
\def\vphot{$v_{\rm phot}$}
\def\ang{~\AA}
\def\syn{{\bf Synow}}
\def\dm15{{$\Delta$}$m_{15}$}
\def\rsi{$R$(Si~II)}
\def\v10{$V_{10}$(Si~II)}
\def\wsi{$W_lambda$(Si~II)}
\def\vdot{\.v(Si~II)}
\def\575{$W(5750)$}
\def\610{$W(6100)$}
\def\tex{$T_{\rm exc}$}
\def\ve{$v_{\rm e}$}

\title {Direct Analysis of Spectra of the Unusual Type~Ib Supernova
2005bf}

\author {Jerod Parrent\altaffilmark{1}, David Branch\altaffilmark{1},
M.~A. Troxel\altaffilmark{1}, D.~Casebeer\altaffilmark{1},
David~J. Jeffery\altaffilmark{1}, W.~Ketchum\altaffilmark{1},
E.~Baron\altaffilmark{1}, F.~J.~D. Serduke\altaffilmark{2}, 
and Alexei~V.~Filippenko\altaffilmark{2}}

\altaffiltext{1} {Homer L. Dodge Department of Physics and Astronomy,
University of Oklahoma, Norman,~OK 73019; e-mail: parrent@nhn.ou.edu,
branch@nhn.ou.edu}

\altaffiltext{2} {Department of Astronomy, University of California,
  Berkeley, CA 94720-3411}

\begin{abstract}

Synthetic spectra generated with the parameterized supernova
synthetic-spectrum code SYNOW are compared to spectra of the unusual
Type~Ib supernova 2005bf.  We confirm the discovery by Folatelli
et~al. (2006) that very early spectra ($\sim$30 days before maximum
light) contain both photospheric-velocity ($\sim$8000~\kms) features
of He~I, Ca~II, and Fe~II, and detached high-velocity
($\sim$14,000~\kms) features of H$\alpha$, Ca~II, and Fe~II.  An early
spectrum of SN~2005bf is an almost perfect match to a
near-maximum-light spectrum of the Type~Ib SN~1999ex.  Although
these two spectra were at very different times with respect to maximum
light (20 days before maximum for SN~2005bf and five days after for
SN~1999ex), they were for similar times after explosion --- about 20
days for SN~2005bf and 24 days for SN~1999ex.  The almost perfect
match clinches the previously suggested identification of H$\alpha$ in
SN~1999ex and supports the proposition that many if not all Type~Ib
supernovae eject a small amount of hydrogen.  The earliest available
spectrum of SN~2005bf resembles a near-maximum-light spectrum of the
Type~Ic SN~1994I.  These two spectra also were at different times with
respect to maximum light (32 days before maximum for SN~2005bf and
four days before for SN~1994I) but at similar times after explosion
--- about eight days for SN~2005bf and 10 days for SN~1994I.  The
resemblance motivates us to consider a reinterpretation of the spectra
of Type~Ic supernovae, involving coexisting photospheric-velocity and
high-velocity features.  The implications of our results for the
geometry of the SN~2005bf ejecta, which has been suggested to be
grossly asymmetric, are briefly discussed.

\end{abstract}


\section{INTRODUCTION}

Among the hydrogen-deficient Type I supernovae, the Type Ib (SN~Ib)
subclass is defined by the absence of strong Si~II and the presence of
He~I lines in the optical spectra. Type Ic supernovae (SNe~Ic) are
broadly similar to SNe~Ib, but lack conspicuous He~I lines. SNe~Ib and
SNe~Ic are generally thought to be core-collapse supernovae whose
progenitor stars lost most, or all, of their hydrogen and helium
envelopes, respectively. [See Filippenko (1997) for a review of
supernova spectral types.]

Supernova 2005bf was an exceptionally interesting Type~Ib event
(Anupama et~al. 2005; Tominaga et~al. 2005; Folatelli et~al. 2006;
hereafter A05, T05, and F06, respectively).  The bolometric light
curve reached an initial maximum about 15 days after explosion but
also a brighter second maximum (which hereafter we refer to as the
maximum, without qualification) about 40 days after explosion.  Such a
double-peaked light curve and such a long rise time to maximum light
had not been observed previously.  Both T05 and F06 invoked a
double-peaked radial distribution of $^{56}$Ni to account for the
light curve, although in other respects the models they discussed were
quite different.

T05 and F06 attributed an absorption feature in very early spectra to
H$\alpha$, forming at the high velocity of about 14,000~\kms, and F06
provided support for the identification by recognizing the presence of
Ca~II and Fe~II features at the same velocity.  A weaker absorption at
the same wavelength in near-maximum-light spectra was attributed by
A05 to H$\alpha$ and/or Si~II \lam6355, while T05 favored Si~II and
F06 only remarked that by maximum light the early feature attributed
to H$\alpha$ had disappeared.  In view of the unusual nature of
SN~2005bf, and of our interest in the issue of whether SNe~Ib and even
SNe~Ic eject some hydrogen during their explosions 
(Filippenko 1992; Branch et~al. 2006, hereafter 
B06), we have carried out a direct analysis of selected spectra of 
SN~2005bf using a revised version (2.0) of the parameterized 
resonance-scattering supernova synthetic-spectrum code, SYNOW.

Spectra of SN~2005bf are displayed and discussed in \S2 and our
analysis of them is presented in \S3.  In \S4, the significance of
relationships between the spectra of SN~2005bf, the Type~Ib SN~1999ex
(Hamuy et al. 2002), and the Type~Ic SN~1994I (Filippenko et al. 1995) 
is explored.  The implications of our results are discussed in \S5.

\section{OBSERVED SPECTRA}

Twelve spectra of SN~2005bf are shown in Figure~1.  Following F06,
epochs are with respect to the date of bolometric maximum light, 9~May
2005 (UT dates are used throughout this paper).  Maximum in the $B$
band occurred about 2.5 days earlier.  The day~$-6, -4$, and $-2$
spectra are from A05.  The previously unpublished day~+5 spectrum was
obtained with the 3-m Shane reflector at Lick Observatory using the
Kast spectrograph (Miller \& Stone 1993); observations and reductions
were similar to those of the day~$+2$ spectrum, which was presented
and described in F06.  The other eight spectra also are from F06.
Mild smoothing has been applied to some of the spectra.  Because in
this paper we are interested only in the spectral features, not in the
shape of the underlying continuum, all observed and synthetic spectra
are ``flattened'' by means of the local normalization prescription of
Jeffery et~al. (2007).


From left to right, the dashed lines in Figure~1 correspond to He~I
\lam5876 blueshifted by 7000 \kms, H$\alpha$ \lam6563 blueshifted by
15,000 \kms, and He~I \lam6678 and \lam7065 blueshifted by 7000 \kms.
He~I absorptions blueshifted by about 7000 \kms\ are clearly present
at all epochs.  [From careful measurements of the wavelength of the
absorption minimum attributed to He~I \lam5876, T05 noticed that the
blueshift increased slightly with time.  This may have been due to the
increasing strength of the feature; all else being equal, stronger
features have higher blueshifts (Jeffery \& Branch 1990).]  The deep
absorption attributed to high-velocity H$\alpha$ at early times
becomes weaker between day~$-20$ and day~$-6$, but a feature persists
at close to the same wavelength as late as day~+5.


\section{ANALYSIS}

For the analysis presented in this paper we have used a revised
version (2.0) of SYNOW\footnote{Version 2.0 will soon be available at
{http://nhn.ou.edu/$\sim$parrent/download.html}.} that is described by
Branch et~al. (2007).  New features employed here are (1) when using
power-law line optical depth profiles, different power-law indices
can be adopted for different ions; (2) a Gaussian line optical depth
profile is now available, so that when detaching an ion from the
photosphere it is not necessary to introduce a discontinuity in the
line optical depth profile; and (3) the output spectra can be
flattened as in Jeffery et~al. (2007).  For this paper the excitation
temperature ($T_{exc}$) has been fixed at a nominal value of 7000~K.
For ions that are not detached from the photosphere a power-law line
optical depth distribution $\tau(v) = \tau_p (v/v_{phot})^{-n}$ is
used, where $v_{phot}$ and $\tau_p$ are the velocity and the line
optical depth at the photosphere.  We use $n =8$ as the default value
of the power-law index, but other values are occasionally adopted to
improve the fits.  For ions that are detached from the photosphere a
Gaussian line optical depth distribution $\tau(v) = \tau_g
\exp^{-((v-v_g)/\sigma_g)^2}$ is used, i.e., the maximum line optical
depth $\tau_g$ occurs at velocity $v_g$.  We refer to ions that are
undetached or only mildly detached as photospheric velocity (PV) ions
and those that are detached at high velocity relative to the
photosphere as HV ions.

In this section we will cover some of the same ground as F06.  (The
necessity of doing so was reinforced for us when we learned that the
expert referee of this paper was not convinced of the HV Fe~II
identification by F06 or by the first version of this paper.)  In this
paper we provide all of the synthetic-spectrum fitting parameters
(unlike F06), and thanks to mild smoothing of the observed spectra and
to the scale on which the fits are published, our fits can be examined
more closely than those of F06.

\subsection {Early Spectra}

Figure~2 shows a close-up view of the multiplet-42 region of Fe~II for
the four early spectra.  The long dashed lines correspond to \lam
4924, \lam5018, and \lam5169, blueshifted by 13,000 \kms.  Absorptions
consistent with these transitions are clearly present in the day $-32, -27$,
and $-24$ spectra.  The short solid lines correspond to the same
transitions blueshifted by 7000 \kms.  The four early spectra have
been compared with SYNOW spectra.  Our interpretation of them is in
good agreement with that of F06, although our results differ in detail
because, for example, the version of SYNOW available to F06 did not
have the Gaussian option.

In Figure~3 the day~$-32$ spectrum is compared with a synthetic
spectrum that has \vphot = 8000 \kms\ and includes lines of PV He~I,
O~I, and Fe~II, as well as HV H~I, Ca~II, and Fe~II.  (The parameters of all
synthetic spectra of this paper are mentioned in the text and/or
listed in Table~1.)  In the synthetic spectrum PV He~I is responsible
for three features and PV O~I \lam7773 is responsible for one.  The
main (but not only) role of PV Fe~II is to produce the dip in the
synthetic spectrum near 5040\ang.  The influence of the HV ions is
greater than that of the PV ions.  The HV Ca~II infrared triplet
produces the deep absorption near 8110\ang, and HV Fe~II produces
absorptions near 4930, 4790, 4710, and 4560\ang, and HV H$\alpha$
produces the deep absorption near 6210\ang.  The $v_g$ values for HV
Ca~II and H~I are 17,000 \kms\ while that of Fe~II is 15,000 \kms.  We
consider the HV Ca~II, H$\alpha$, and Fe~II identifications to be
definite.  Figure~4 is like Figure~3 except that HV Fe~II has been
removed.  We know of no way to restore the good fit of Figure~3 without
using HV Fe~II.

In Figure~5 the day~$-27$ spectrum is compared with a synthetic
spectrum that is much like that of Figure~3 for day~$-32$, but in
Figure~5 \vphot = 7000 \kms\ and PV Ca~II is introduced.  (The
difference between \vphot = 8000 \kms\ for day~$-32$ and \vphot = 7000
\kms\ for day~$-27$ is within our fitting uncertainties.)  The HV and
PV Ca~II parameters are chosen to produce a reasonable fit to the
P-Cygni profile extending from about 8000\ang\ to 8800\ang.  The main
changes between the synthetic spectra for day~$-32$ and day~$-27$ are
that $\tau_p$(He~I) has increased by a factor of 1.5, $\tau_g$(HV
Ca~II) has decreased by a factor of 6.25, and PV Ca~II has been
introduced with $\tau_p = 4$.  T05 used a different
synthetic-spectrum code, the Mazzali-Lucy Monte Carlo code (Mazzali
2000), to fit a day~$-26$ spectrum.  They attributed the 6210\ang\
absorption to a blend of HV H$\alpha$ and PV Si~II, and they did not
explicitly\footnote{In principle, since T05 input density and
abundance, not line optical depth, their synthetic spectra could have
contained HV Ca~II and Fe~II features.} introduce HV Ca~II and Fe~II
features. (Their day~$-26$ spectrum did not extend far enough to the
red to show the HV Ca~II absorption at 8000\ang.)


Between days~$-27$ and $-24$ there are no major changes in our
synthetic-spectrum parameters (see Table~1).  In Figure~6 the
day~$-20$ spectrum is compared with a synthetic spectrum that has
\vphot = 8000 \kms.  The major changes compared to day~$-27$ are that
for day~$-20$ HV Fe~II is not used at all, $\tau_p$(PV Ca~II) has
increased by a factor of 12.5, and $\tau_p$(PV Fe~II) has increased by
a factor of 10.


Significant spectral evolution occurred between days~$-32$ and
$-20$, especially between days $-24$ and $-20$.  The day~$-32$ 
spectrum is strongly influenced by HV
features, but PV features also are present.  By day~$-20$, the
spectrum is mainly PV, the only HV feature being H$\alpha$.  From
day~$-32$ to day~$-20$, $\tau_p$(He~I) increased by a factor of two,
suggestive of increasing nonthermal excitation (T05), while the
fitting parameters for H$\alpha$ varied only mildly.

\subsection{Maximum Light}

The five spectra of Figure~1 from day~$-6$ to day~+5 are rather
similar, so we consider only day~+2 here.  In Figure~7 the day~$+2$
spectrum is compared with a synthetic spectrum that has \vphot = 7000
\kms\ and includes lines of PV He~I, O~I, Ca~II, and Fe~II, as well as HV
H~I.  A05 noted that in the day~$-6$ spectrum the Fe~II absorptions
were more blueshifted than the He~I absorptions.  We find the same for
the day~+2 spectrum: in the synthetic spectrum of Figure~7 Fe~II lines
are mildly detached,\footnote{Strictly speaking, an ion is not
detached if its line optical depth reaches a maximum value above the
photosphere but is not negligible at the photosphere, as is the case
here.  Nevertheless, for brevity we refer to such cases as detached.}
at 8000 \kms, while the He~I lines are not.

T05 found that in a day~$-5$ spectrum Si~II was satisfactory for the
absorption that we attribute to H$\alpha$.  Figure~8, which is like
Figure~7 except that HV H~I has been replaced by PV Si~II ($\tau_p$=1,
$n$=8), shows that for us the Si~II absorption has its usual problem
in SN~Ib/c spectra: it is too blue to account for the observed
absorption on its own.  The synthetic spectrum of T05 has \vphot =
4600 \kms, compared to our value of 7000 \kms, and from their Figure~4
it appears that many of the synthetic absorptions are insufficiently
blueshifted.  This may account for their finding that Si~II is
satisfactory.

\subsection{Postmaximum}

The latest three spectra of Figure~1 are similar, so we consider only
day~+21.  In Figure~9 the day~+21 spectrum is compared with a
synthetic spectrum that has \vphot = 5000 \kms\ and includes the same
lines as in Figure~7 except that HV H~I is not used --- the synthetic
spectrum is entirely PV.  The observed absorptions near 5460 and
5590\ang\ could be fitted reasonably well with PV Sc~II, but then
Sc~II \lam4247 would produce a deep unwanted absorption near
4100\ang. Part of the observed depression extending from 6290 to
6480\ang\ could be fit by introducing PV H$\alpha$ but the
identification would not be convincing.


\section{RELATIONSHIP TO SPECTRA OF SN~1999ex AND SN~1994I}

F06 noted that the day~$-20$ spectrum of SN~2005bf and a day~+5
spectrum of the Type~Ib SN~1999ex (Hamuy et al. 2002) were similar, in
spite of the very different epochs with respect to maximum light.  In
Figure~10 the same two spectra are compared directly in a plot of
linear flux and with both spectra flattened (F06 presented a log-flux
plot of unflattened spectra).  Figure~10 shows that the two spectra
are almost identical!  They are so similar that it is not plausible
that two different interpretations apply.  Only minor tweaking of the
synthetic spectrum for day~$-20$ of SN~2005bf (Figure~6), such as
increasing the optical depth of HV Ca~II, would produce an equally
good synthetic spectrum for day~+5 of SN~1999ex.  As discussed above,
the day~$-$20 spectrum of SN~2005bf is dominated by PV features but
also includes HV H$\alpha$.  The spectrum of SN~1999ex was interpreted
in a similar way by B06, but the HV H$\alpha$ identification was
perhaps not entirely convincing.  Figure~10 clinches the HV H$\alpha$
identification in SN~1999ex.


To account for the fact that based on very early spectra, SN~2005bf
was initially classified as a Type~Ic (Morell et~al. 2005; Modjaz,
Kirshner, \& Challis 2005), F06 also noted that to some extent the
day~$-32$ spectrum of SN~2005bf resembled a day~$-5$ spectrum of the
Type~Ic SN~1994I (Filippenko et al. 1995).  This leads us to ask
whether the day~$-32$ spectrum of SN~2005bf has something to teach us
about how to interpret the near-maximum spectra of SN~1994I.
Figure~11 is like Figure~10, but for the day~$-32$ spectrum of
SN~2005bf and a day~$-4$ spectrum of SN~1994I.  The resemblance in
Figure~11 is not as striking as in Figure~10, but it is sufficient to
suggest that previous interpretations of the SN~1994I spectrum should
be reconsidered.  Millard et~al. (1999), using the SYNOW code to
interpret the day~$-4$ spectrum of SN~1994I, did not consider the
possibility that the spectrum was a composite HV and PV spectrum;
therefore in order to fit the Ca~II IR triplet feature they had to use
a high value of \vphot = 17,500 \kms.  This led to difficulties in
accounting for some of the other features. Millard et~al.  attributed
the feature produced by He~I \lam5876 in SN~2005bf to Na~I (with an
imposed maximum velocity) in SN~1994I, and the feature produced by HV
H$\alpha$ in SN~2005bf as a blend of Si~II (with an imposed maximum
velocity) and (detached) C~II in SN~1994I.  

Here we briefly consider a reinterpretation of the day~$-4$ spectrum
of SN~1994I.  Using the same ions as for the day~$-32$ spectrum of
SN~2005bf, we varied the SYNOW input parameters and obtained the fit
shown in Figure~12.  The synthetic spectrum has \vphot = 12,000 \kms\
(compared to 8000 \kms\ for SN~2005bf) and the values of $v_g$ for HV
H~I, HV~Ca~II, and HV~Fe~II are 23,000, 19,000, and 18,000 \kms,
respectively (compared to 17,000, 17,000, and 15,000 \kms\ for
SN~2005bf).  The fit is encouraging (although to fit the Ca~II
infrared triplet we have had to make Ca~II H\&K too strong) and seems
at least as plausible as that of Millard et~al. (1999), as well as
that of Sauer \& Mazzali (2006) who also attributed features in
SN~1994I to Na~I and to a blend of Si~II and C~II (and Ne~I).  A more
thorough reconsideration of the early spectra of SNe~Ic is deferred to
a separate paper.

B06 discussed the possible presence of HV H$\alpha$ and PV He~I in
spectra of SN~1994I, as well as the possible presence of PV H$\alpha$
as would be required by the tentative identification (Filippenko 1988,
1992) of centrally peaked H$\alpha$ emission in SN~1994I and other
SNe~Ic.  (Centrally peaked emission cannot be produced by highly
detached hydrogen.)  Figure~13 is like Figure~12 but with PV H$\alpha$
included (with $\tau_p$ = 0.6, $n$ = 8).  PV H$\alpha$ certainly does
no harm; in fact, the fit is somewhat improved.  It may be that PV
H$\alpha$ actually is in net emission in SN~1994I, as it is in SNe~II,
rather than a resonance-scattering profile as given by SYNOW; net
emission in the synthetic spectrum would improve the fit near
6563\ang. The presence of PV H$\alpha$ in SNe~Ic remains a
possibility.


\section{DISCUSSION}

Our interpretation of the very early spectra, including the
identification of HV H$\alpha$, is in good agreement with that of F06.
In addition, we favor the H$\alpha$ identification as late as day~+2,
and perhaps even day~+5.
Wang \& Baade (2005) reported that an unpublished spectrum obtained on
day~$-9$ (presumably a very high signal-to-noise spectrum, because
spectropolarimetry was obtained) appeared to contain H$\beta$ and
H$\gamma$ absorptions.  This increases our confidence in the H$\alpha$
identification\footnote{Soderberg et~al. (2005) reported that at
day~175 the nebular spectrum was dominated by strong, broad (FWHM
$\sim$3400 \kms) H$\alpha$ emission, but this emission may have been
circumstellar rather than from low-velocity ejecta.} at day~+2.


F06 discussed models for SN~2005bf.  Although the numerical results
they presented were based on spherical symmetry, they envisaged a
grossly asymmetric explosion having many features in common with the
supernovae that are associated with gamma-ray bursts, for which the
most popular model is the collapsar model (Woosley \& Bloom 2006, and
references therein).  In SN~2005bf, according to F06, a collapsar
launched relativistic jets that drove a small fraction of the ejected
matter into high-velocity ($v \ga 14,000$ \kms) bipolar flows that
contained about 0.1~M$_\odot$ per pole of $^{56}$Ni, were optically
thick at early times, and produced the first light-curve peak.  When
the bipolar ejecta became optically thin, the underlying low-velocity
ejecta, primarily equatorial and containing most of the mass and
$^{56}$Ni, powered the main light-curve peak.  It should be noted,
though, that the presence of PV features in the early spectra shows
that the bipolar flows were {\sl not} optically thick at the time of
the first peak.  Another issue is the distribution of hydrogen with
respect to velocity.  The progenitor star contained a surface layer of
hydrogen.  If the bipolar flows are of sufficiently wide opening
angle, it may be that a pole-on observer sees only HV hydrogen in
absorption, not PV hydrogen.  But an equator-on observer would see PV
hydrogen in absorption.  Equator-on is statistically more likely than
pole-on, yet it is not clear that any SN~Ib/c has been seen to have
PV, but not HV, hydrogen.  Thus the F06 interpretation of SN~2005bf
requires that SN~2005bf-like events are quite uncommon; otherwise,
SNe~Ib/c with PV but not HV hydrogen in absorption should be found.

Like F06, T05 argued that the double-peaked light curve required a
double-peaked radial distribution of $^{56}$Ni, but T05 invoked jets
that were not sufficiently energetic to reach the bottom of the helium
layer, at $ v \simeq 6000$ \kms; these jets provided a small amount of
$^{56}$Ni at intermediate velocities ($ 3900 \la v \la 5400$ \kms) to
power the first peak.\footnote{The referee asks us to mention that
both F06 and T05 had to make the unphysical assumption that the
gamma-ray opacity abruptly decreased in order to simulate the rapid
decline of the light curve after the second peak.}  Based on several
perceived similarities, T05 suggested that SN~2005bf may have been a
Cas~A-like event.  The T05 model is not necessarily highly
asymmetric,\footnote{But in order to fit the rapid post-maximum
decline of the light curve, both T05 and F06 had to assume a rapid drop
in the gamma-ray opacity, which could be a consequence of strong
asymmetry.} so all of the hydrogen could be ejected at high velocity,
and the issue of seeing SNe~Ib/c with PV but not HV hydrogen does not
necessarily arise.


In view of the almost perfect resemblance of the day~$+5$ spectrum of
SN~1999ex and the day~$-20$ spectrum of SN~2005bf (Fig.~8), we regard
the identification of H$\alpha$ in SN~1999ex as definite.  This
supports the proposition that 
most
SNe~Ib eject some
hydrogen (Deng et~al. 2000; Branch et~al. 2002; Elmhamdi et~al. 2006).
To our knowledge, there is no published, clear explanation 
why hydrogen should 
generally not 
be entirely removed from the progenitors of supernovae
that develop conspicuous He~I lines.


The resemblance of the day~$+5$ spectrum of SN~1999ex and the
day~$-20$ spectrum of SN~2005bf means that at these epochs the
conditions near the photosphere --- composition, density structure,
and temperature --- were very similar.  It is interesting to note (a)
that for SN~2005bf day~$-20$ corresponds to about day~$+5$ with
respect to the {\sl initial} maximum, and (b) that these epochs
correspond to similar times after explosion --- 20 days for SN~2005bf
and 24 days for SN~1999ex [assuming a rise time for SN~1999ex of 19
days (Richardson, Branch, \& Baron 2006)].  But although the
conditions at the photosphere were similar at these epochs, what was
to follow --- to be determined by what was still {\sl beneath} the
photosphere --- would be very different.


Similarly, the times with respect to explosion for the day~$-4$
spectrum of SN~1994I and the day~$-32$ spectrum of SN~2005bf (Fig.~10)
are not very different --- about eight days for SN~2005bf and 10 days
for SN~1994I [assuming a rise time for SN~1994I of 14 days (Richardson
et~al 2006)].  The similarities between these two spectra raise our
suspicion that SN~1994I ejected hydrogen.  If so, then
most or all
other ordinary
SNe~Ic also do (see B06), and they are not explosions of bare
carbon-oxygen cores as they usually are modelled.  If SN~1994I did
{\sl not} eject hydrogen, then the spectroscopic coincidences between
SNe~Ib that do eject hydrogen and SNe~Ic that do not eject hydrogen
are even more striking than they appeared to B06.

We are grateful to G. Anupama and G. Folatelli for providing spectra.
We also thank the staffs at the Lick and Keck Observatories for
this assistance.
This work has been supported by NSF grants AST--0204771,
AST--0506028, and AST--0607485, as well as by NASA LTSA grant NNG04GD36G.

\clearpage     


\begin{figure}
\includegraphics[width=.8\textwidth,angle=0]{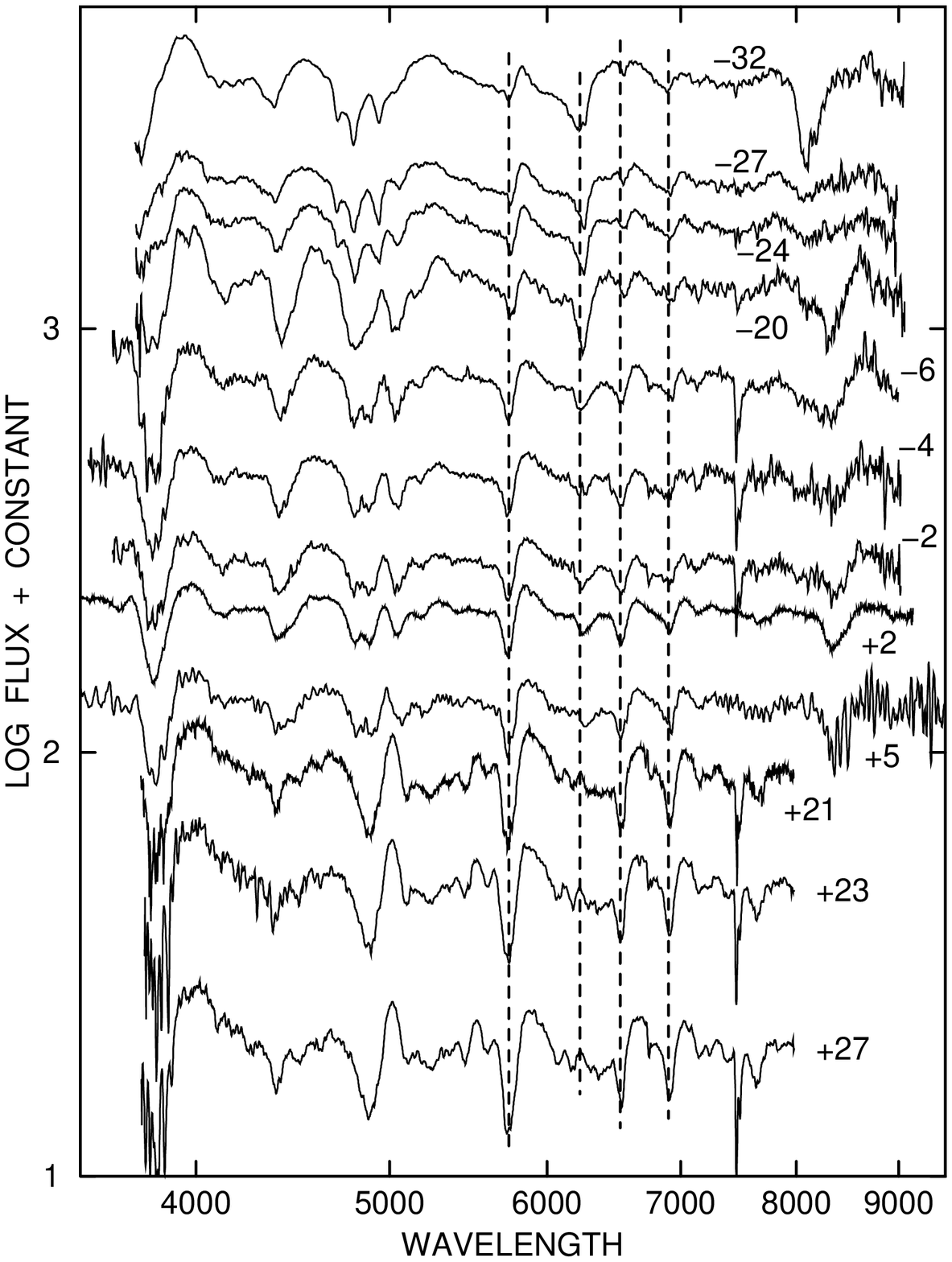}
\caption{Spectra of SN~2005bf from A05 and F06, and a previously
unpublished spectrum for day~+5.  The spectra have been corrected for
$cz = 5496$ \kms\ (F06).  The narrow absorption near 7460\ang\ is
telluric and vertical shifts are arbitrary.  The spectra have been
flattened by means of the local normalization prescription of Jeffery
et~al. (2007).  The {\sl dashed lines} are discussed in the text.}
\end{figure}


\begin{figure}
\includegraphics[width=.8\textwidth,angle=270]{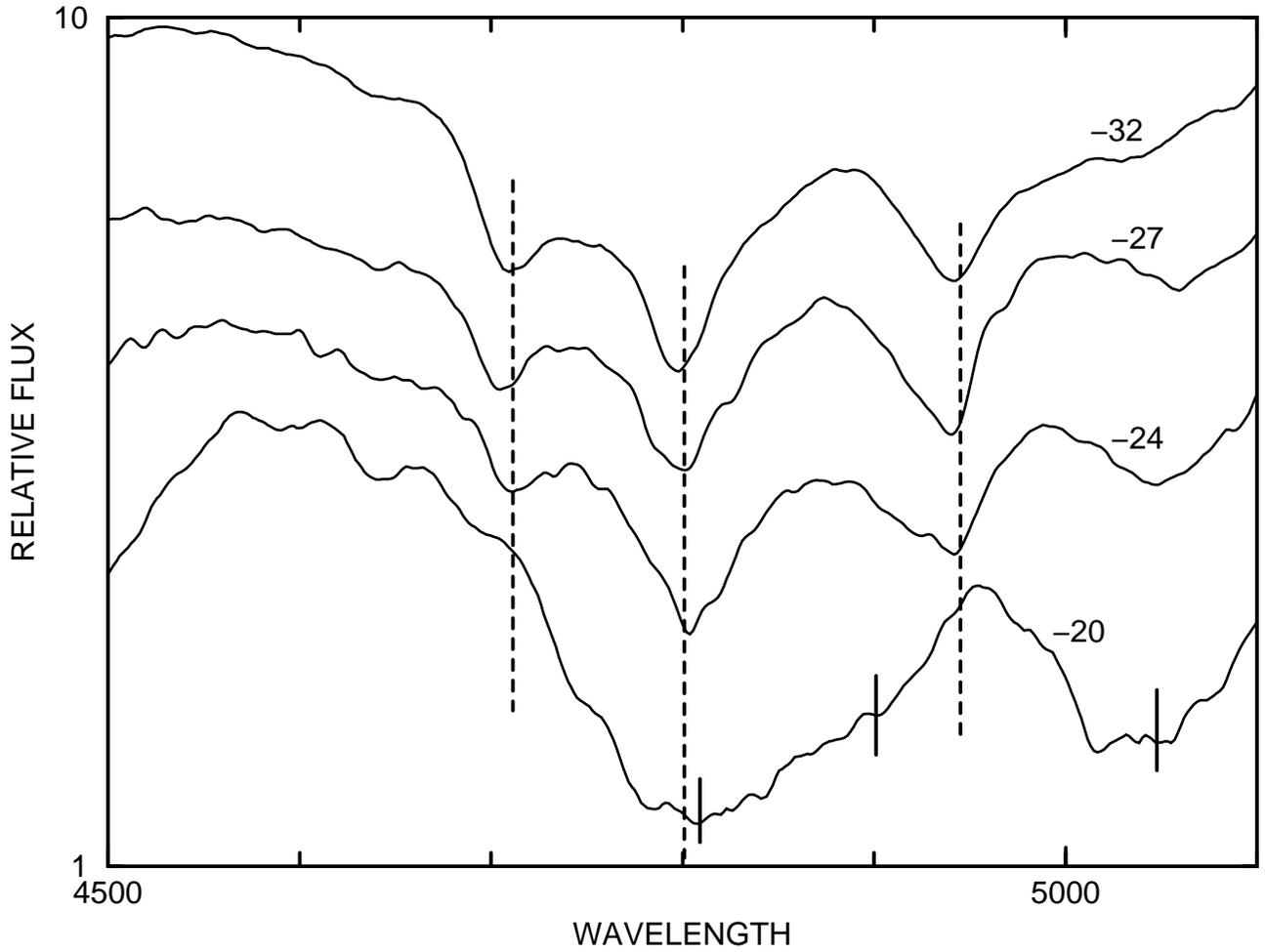}
\caption{A close-up view of the multiplet-42 region of Fe~II in the four
earliest spectra of SN~2005bf, from F06.  The long {\sl dashed lines}
correspond to Fe~II \lam4924, \lam5018, and \lam5169 blueshifted by
13,000 \kms, and the short {\sl solid lines} correspond to the same
lines blueshifted by 7000 \kms.}
\end{figure}

\begin{figure}
\includegraphics[width=.8\textwidth,angle=270]{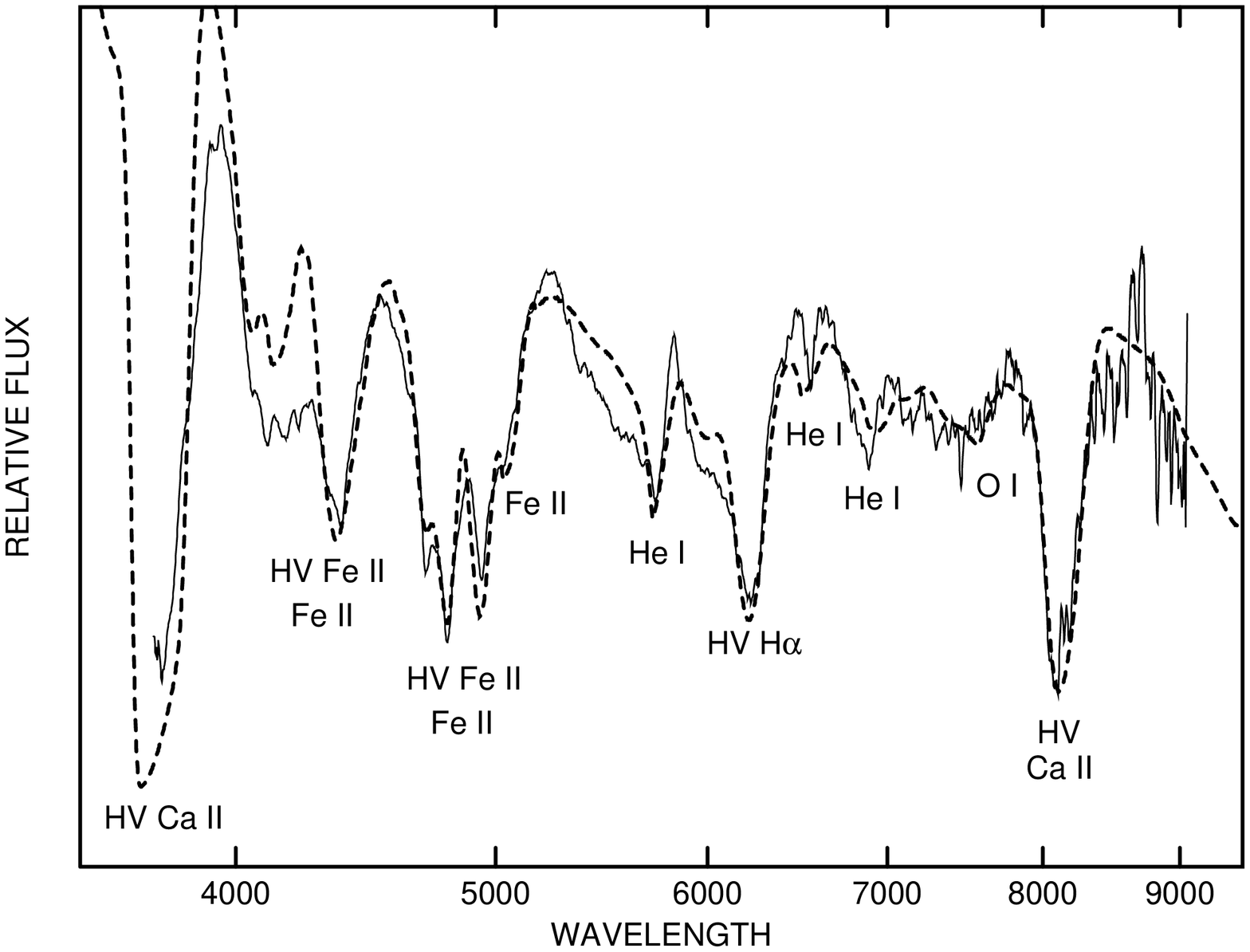}
\caption{The day~$-32$ spectrum of SN~2005bf ({\sl solid line}) is
  compared with a synthetic spectrum ({\sl dashed line}). }
\end{figure}


\begin{figure}
\includegraphics[width=.8\textwidth,angle=270]{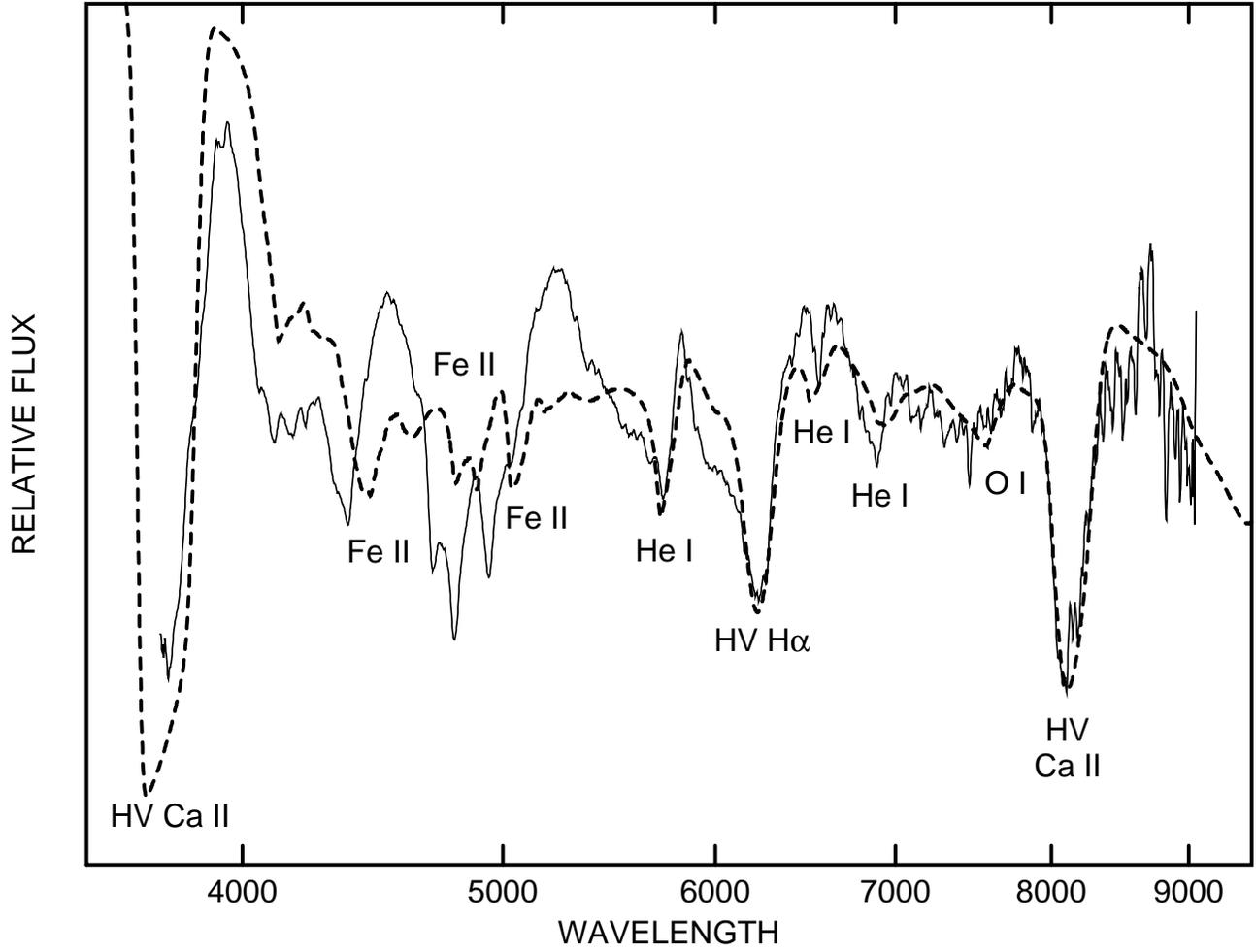}
\caption{Like Figure~3 except that HV Fe~II lines have been removed.}
\end{figure}


\begin{figure}
\includegraphics[width=.8\textwidth,angle=270]{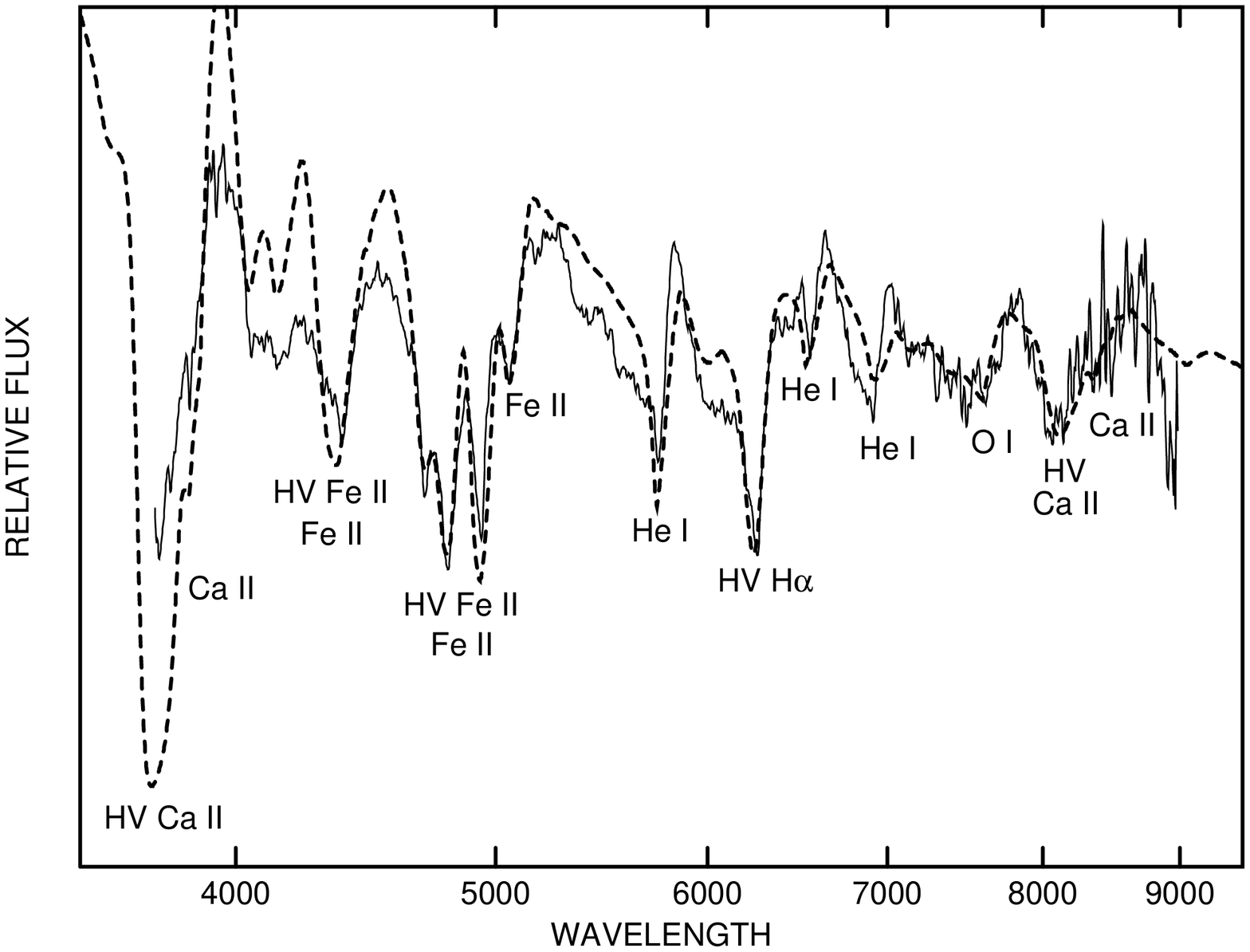}
\caption{The day~$-27$ spectrum of SN~2005bf ({\sl solid line}) is
  compared with a synthetic spectrum ({\sl dashed line}). }
\end{figure}

\begin{figure}
\includegraphics[width=.8\textwidth,angle=270]{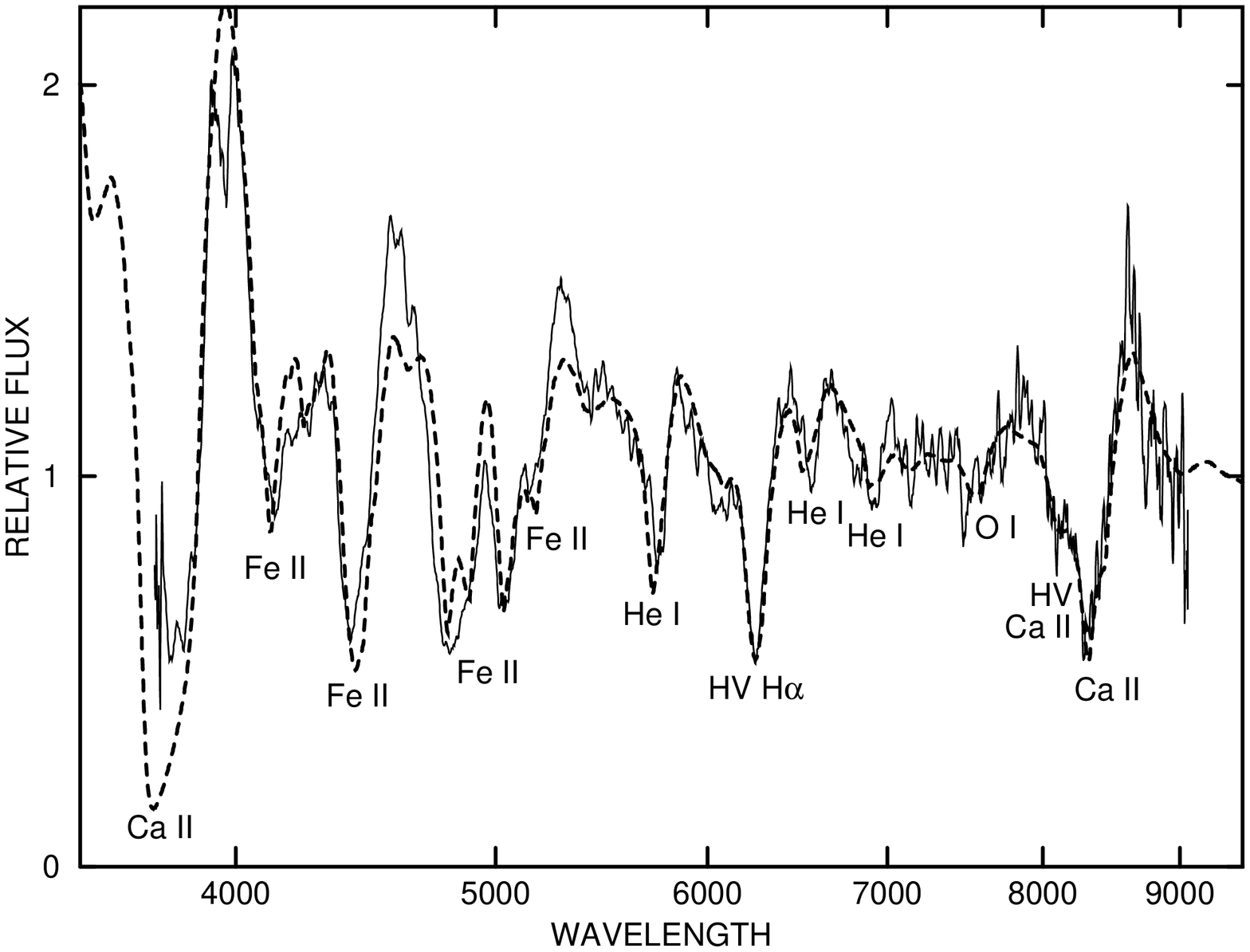}
\caption{The day~$-20$ spectrum of SN~2005bf ({\sl solid line}) is
  compared with a synthetic spectrum ({\sl dashed line}). }
\end{figure}

\begin{figure}
\includegraphics[width=.8\textwidth,angle=270]{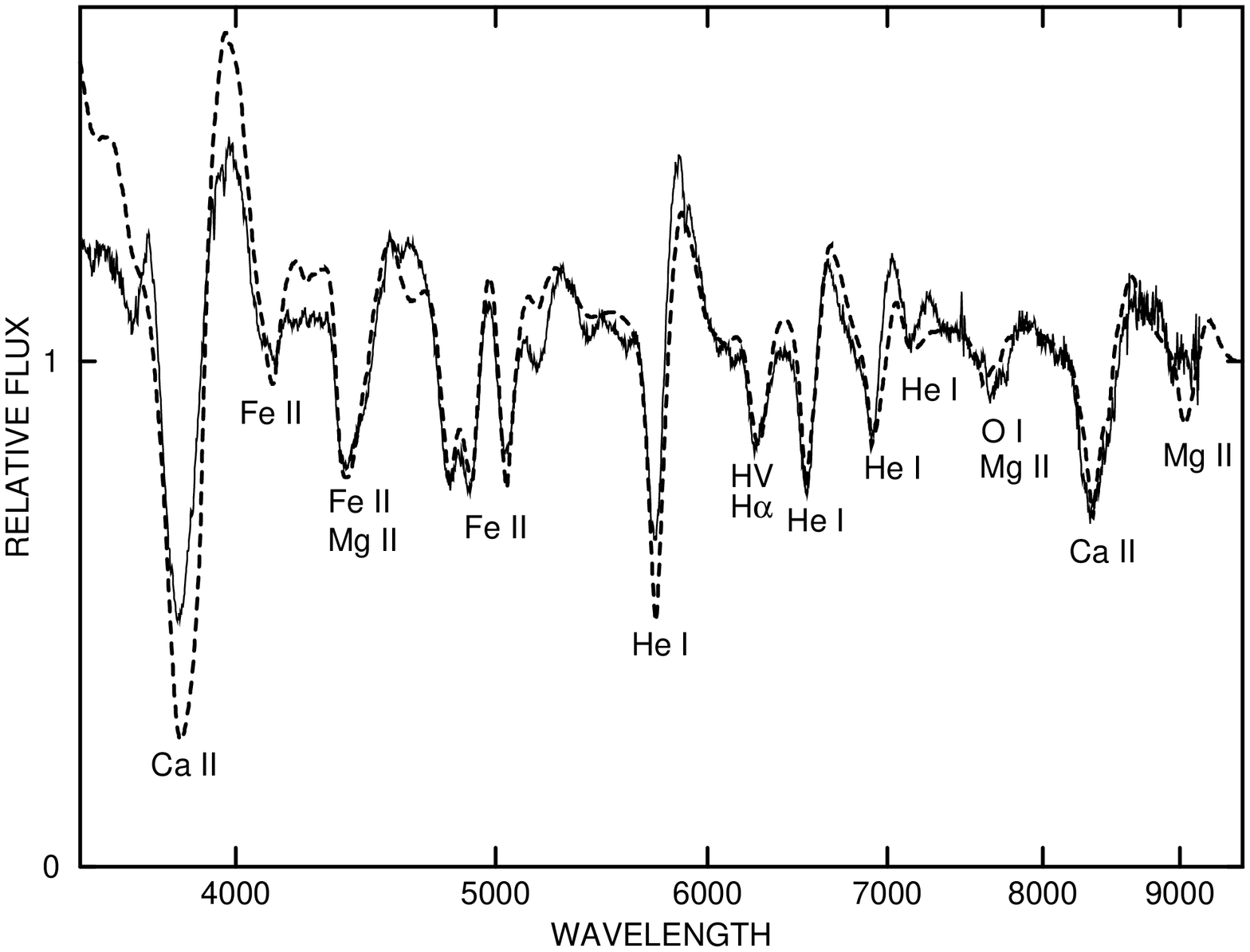}
\caption{The day~+2 spectrum of SN~2005bf ({\sl solid line}) is
  compared with a synthetic spectrum ({\sl dashed line}). }
\end{figure}

\begin{figure}
\includegraphics[width=.8\textwidth,angle=270]{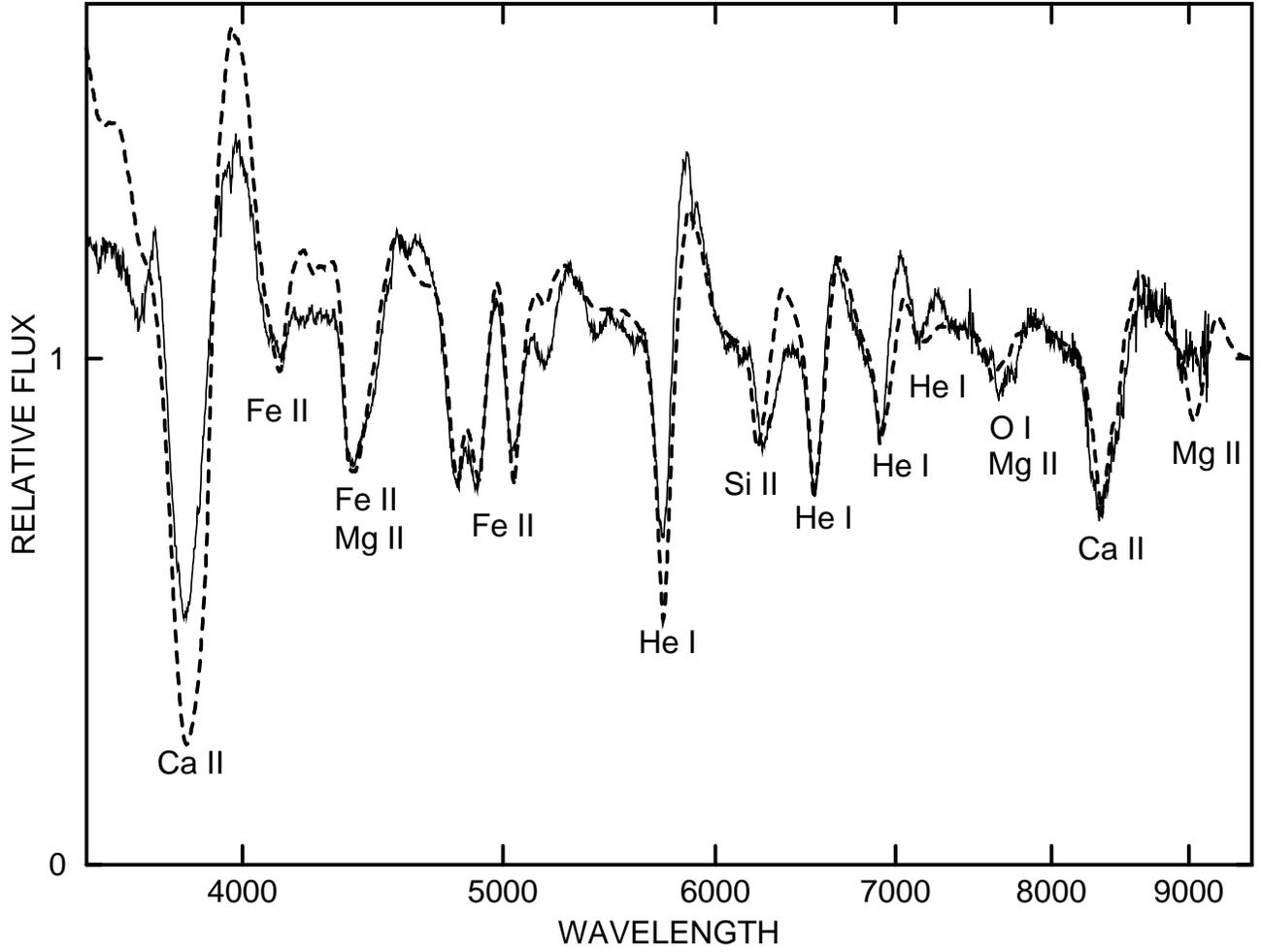}
\caption{Like Figure~5 but with HV hydrogen replaced by PV Si~II.}
\end{figure}

\begin{figure}
\includegraphics[width=.8\textwidth,angle=270]{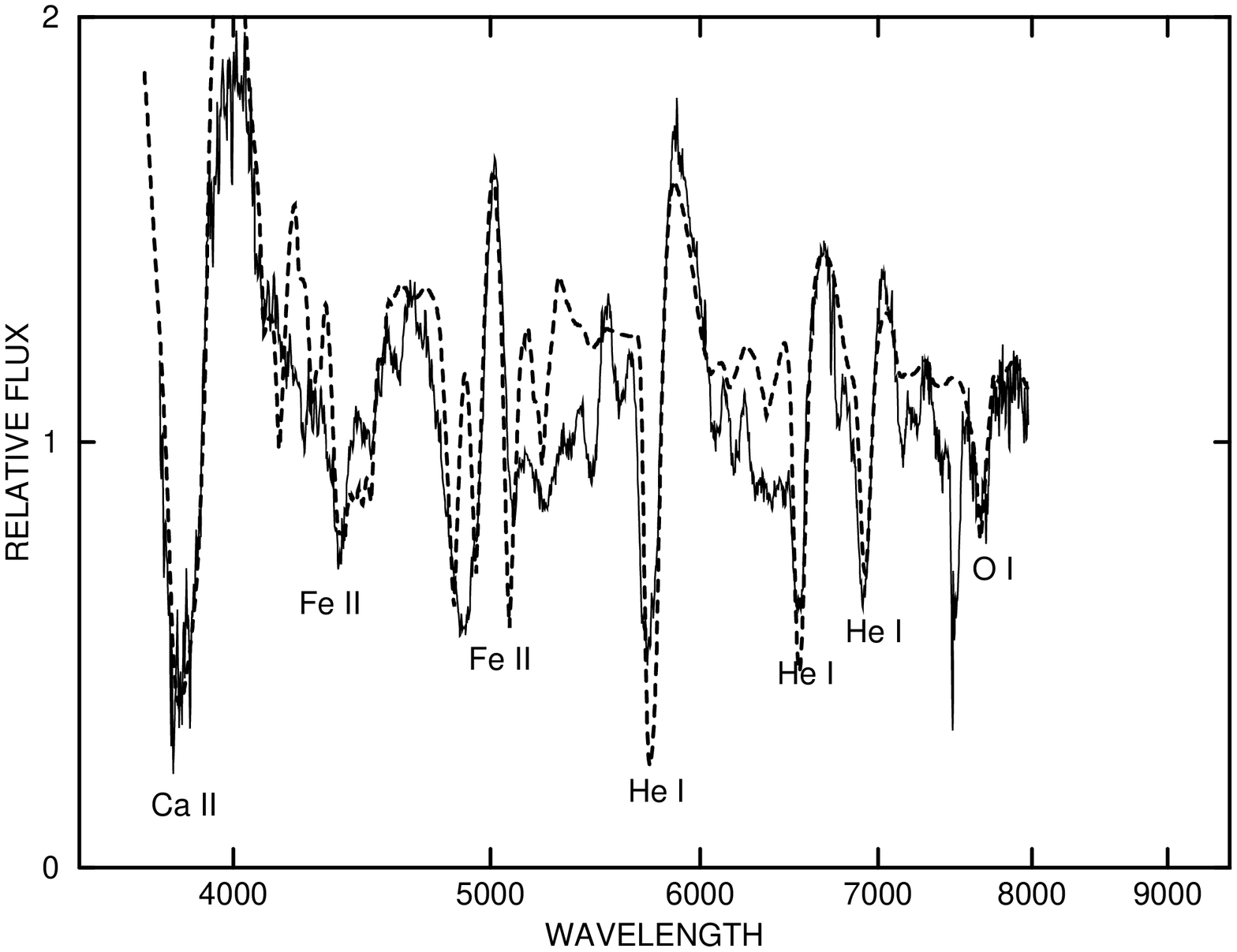}
\caption{The day~+21 spectrum of SN~2005bf ({\sl solid line}) is
  compared with a synthetic spectrum ({\sl dashed line}). }
\end{figure}

\begin{figure}
\includegraphics[width=.8\textwidth,angle=270]{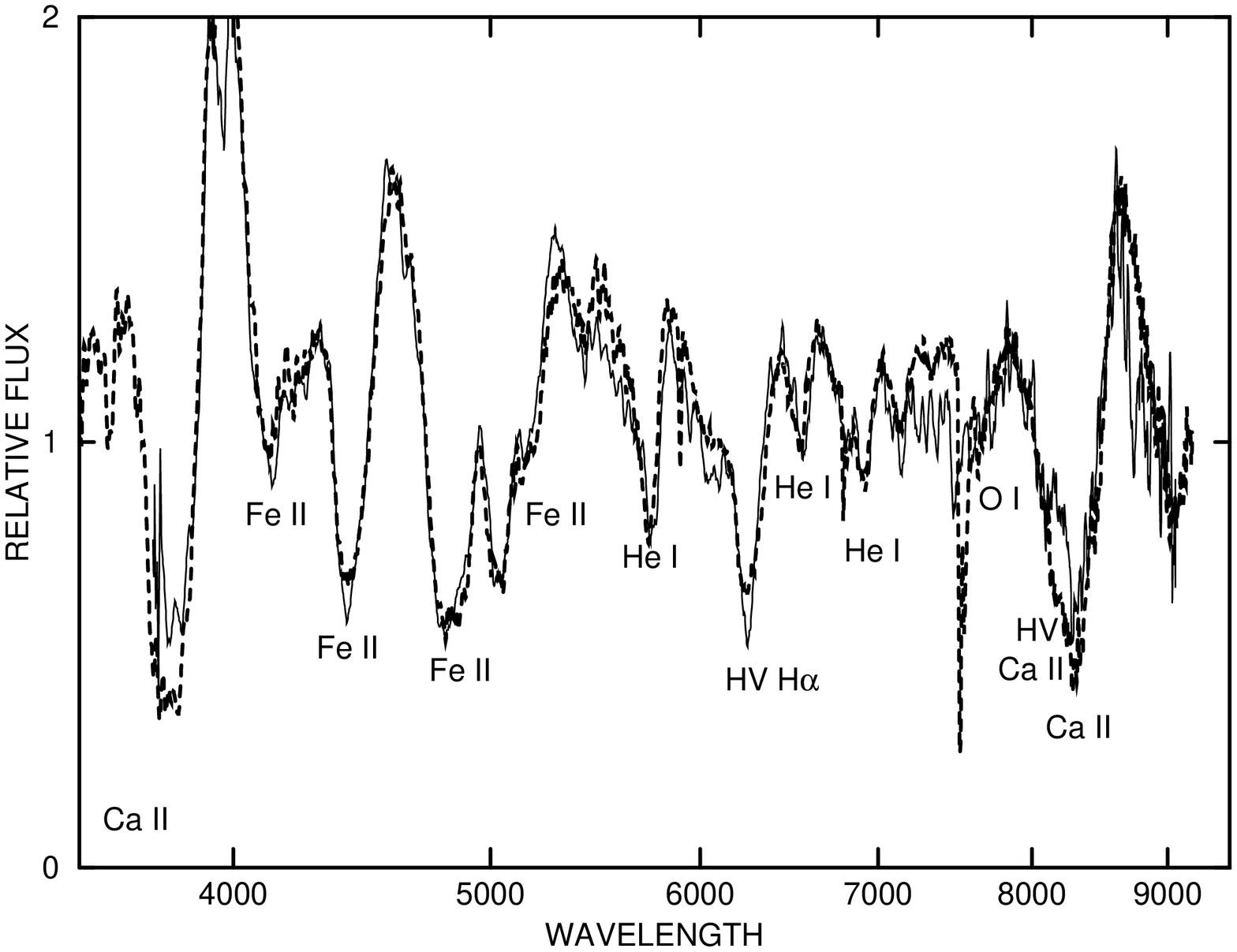}
\caption{A day~+5 spectrum of SN~1999ex ({\sl dashed line}) is
  compared with the day~$-20$ spectrum of SN~2005bf ({\sl solid line}). }
\end{figure}

\begin{figure}
\includegraphics[width=.8\textwidth,angle=270]{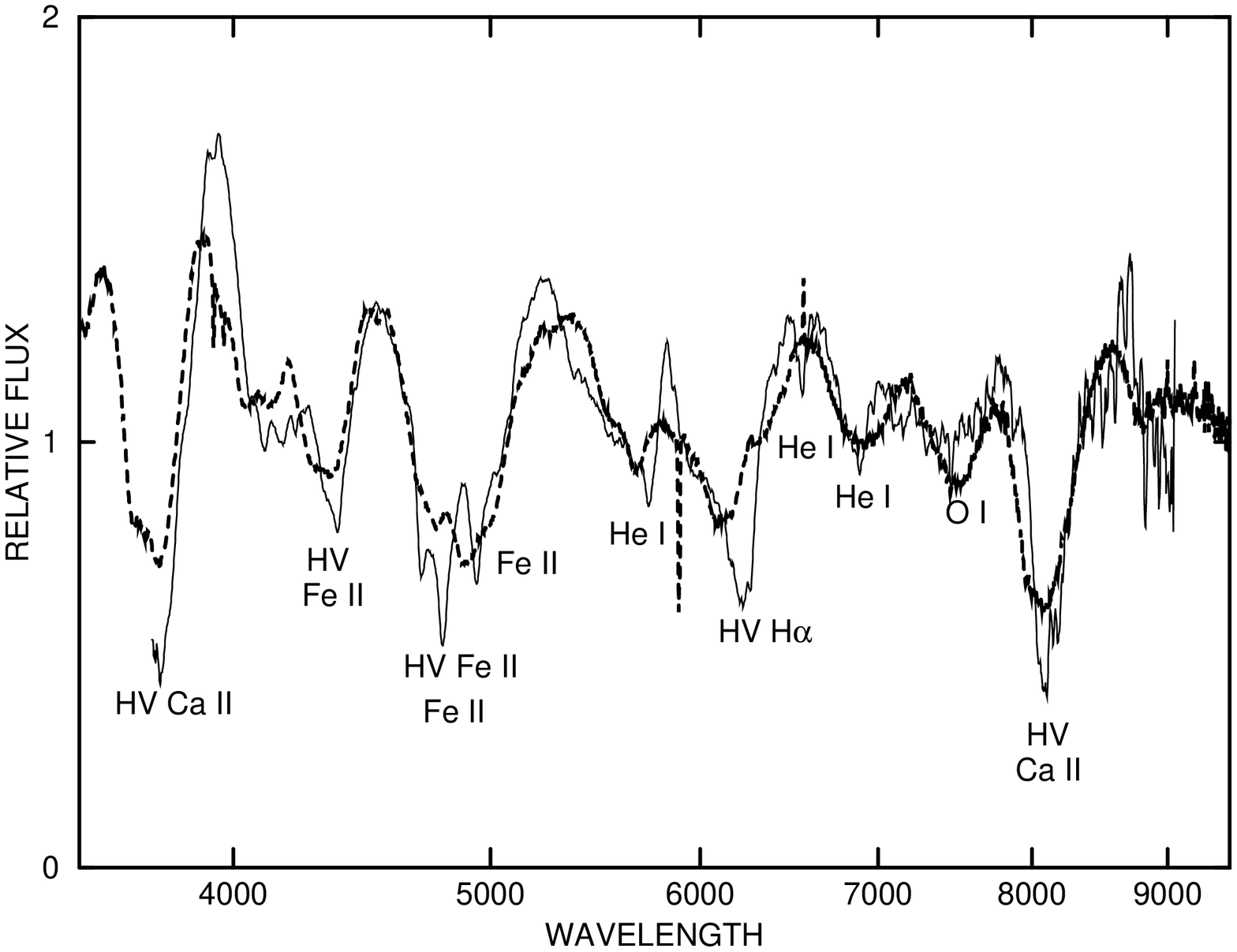}
\caption{A day~$-4$ spectrum of SN~1994I ({\sl dashed line}) is
  compared with the day~$-32$ spectrum of SN~2005bf ({\sl solid
  line}). }
\end{figure}


\begin{figure}
\includegraphics[width=.8\textwidth,angle=270]{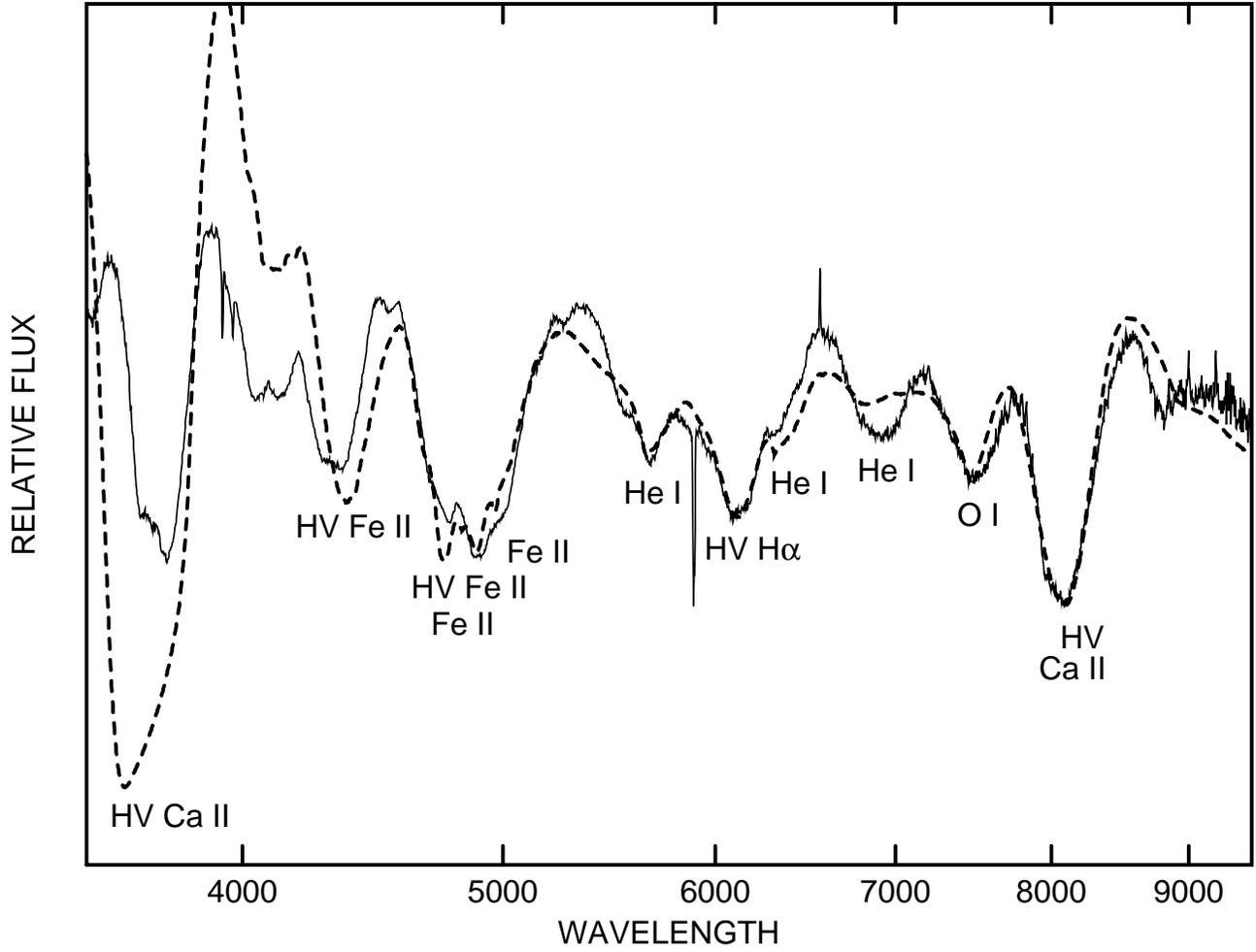}
\caption{The day~$-4$ spectrum of SN~1994I ({\sl solid line}) is
  compared with a synthetic spectrum ({\sl dashed line}). }
\end{figure}


\begin{figure}
\includegraphics[width=.8\textwidth,angle=270]{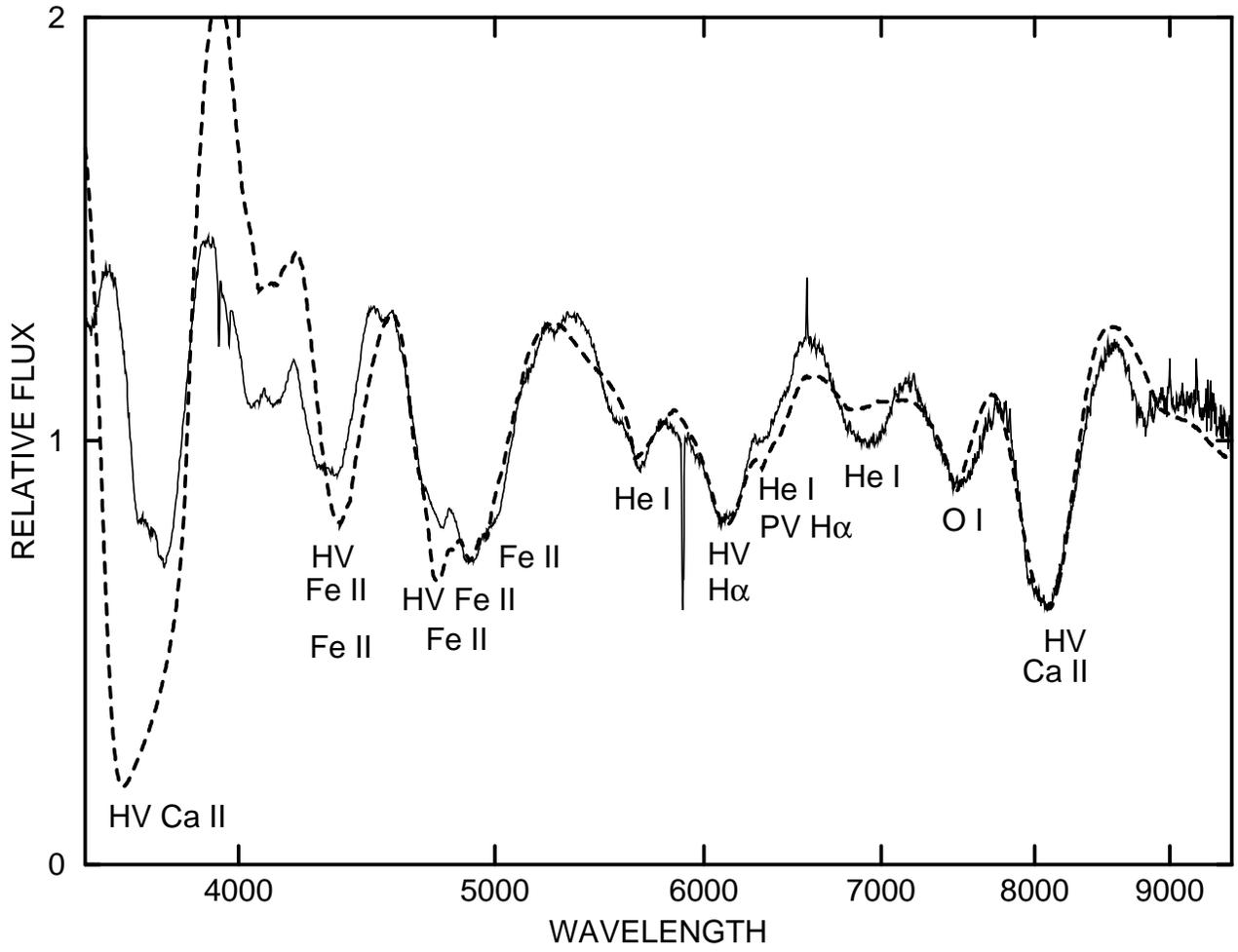}
\caption{Like Figure~12 except that PV H~I is included.}
\end{figure}

\clearpage


\begin{deluxetable}{lccccccccc}
\tablenum{1}
\setlength{\tabcolsep}{4pt}

\tablecaption{Fitting Parameters for Synthetic Spectra$^a$}

\tablehead{\colhead{Parameter} & \colhead{Fig. 2} & \colhead{Fig. 3} &
\colhead{Fig. 4} & \colhead{Fig. 5}& \colhead{Fig. 6} &
\colhead{Fig. 7} & \colhead{Fig. 12}& \colhead{Fig. 13}}


\startdata

\vphot\  & 8000 & 7000 &  8000 & 7000 & 7000 & 5000 &
12,000 & 12,000\\

$\tau_p$(PV H~I) & 0 & 0 & 0 & 0 &0 & 0 & 0 &0.6\\

n(PV H~I) & ... & ... & ... & ... & ... & ... & ... &8 \\

$\tau_g$(HV H~I) & 0.7 & 0.6 & 0.8 &  0.25 &0 & 0 & 0.25 &0.25\\ 

$v_g$(HV H~I) & 17,000 & 16,000 & 16,000 & 15,000 & ...&
... & 23,000 &23,000\\

$\sigma_g$(HV H~I) & 3000 & 2000 & 2000 & 2000 & ...& ... & 4000&4000\\

$\tau_p$(PV He~I) & 0.8 & 1.2 & 1.6 & 4 &4 & 0 & 0.3 &0.3\\

n(PV He~I) & 8 & 8 & 8 & 8 & 8 & ... & 8 &8 \\

$\tau_g$(PV He~I) & 0 & 0 & 0 &0 &0 & 2.5 & 0& 0\\

$v_g$(PV He~I)  & ... & ... & ... & ... &... &
7000 & ... & ...\\

$\sigma_g$(PV He~I)& ... & ... & ... & ... & ... & 1000 & ... & ...\\

$\tau_p$(PV O~I) & 0.1& 0.12 & 0.1  & 0.2 &0.2 & 1 & 0.3 &0.3\\

n(PV O~I) & 2 & 2 & 2 & 8&8 & 8 & 4& 4\\

$\tau_p$(PV Mg~II) &0& 0 & 0 & 0.5 &0.5 & 2 & 0 &0\\

n(PV Mg~II) & ... & ... & ... & 8 &8 & 8 & ... & ...\\

$\tau_p$(PV Si~II) &0& 0 & 0 & 0 & 1 & 0 & 0 &0\\

n(PV Si~II) & ... & ... & ... & ... &8 & ... & ... & ...\\

$\tau_p$(PV Ca~II) & 0 & 12 &150 &  40&40 & 40 & 0&0\\

n(PV Ca~II) & ... & 8 & 8 & 6 &6 & 4 & ...& ...\\

$\tau_g$(HV Ca~II) & 25 & 4 &4.5  & 0 &0 & 0 & 12 &12\\

$v_g$(HV Ca~II) & 17,000 & 17,000 & 17,000 & ... &...&
... & 19,000&19,000\\

$\sigma_g$(HV Ca~II) & 2500 & 2500 & 2000 & ... &...& ... & 6000&6000\\

$\tau_p$(PV Fe~II) & 1.5 & 1 & 10 & 0 &0 & 10 & 2&2\\

n(PV Fe~II) & 8 & 8 & 8 & ... &...& 8 & 8&8\\

$\tau_g$(PV Fe~II) & 0 & 0 & 0 & 0.8 &0.8 & 0 & 0&0\\

$v_g$(PV Fe~II) & ... & ... & ... & 8000 &8000&
... & ...&...\\

$\sigma_g$(PV Fe~II) & ... & ... & ... & 2000 &2000& ... & ...&...\\

$\tau_g$(HV Fe~II) & 1 & 0.9 &0  & 0 &0 & 0 &0.5&0.5\\

$v_g$(HV Fe~II) & 15,000 & 15,000 & ... & ... &....&
... & 18,000&18,000\\

$\sigma_g$(HV Fe~II) & 2000 & 2000 & ... &  ... &...& ... & 2000&2000\\

\enddata

\tablenotetext{a}{The units of $v_{phot}, v_g$, and $\sigma_g$ are
\kms.}

\end{deluxetable}

\end{document}